\documentclass[12pt, a4paper]{article}

\pdfoutput=1

\usepackage{amsmath}
\allowdisplaybreaks
\usepackage{amsfonts}
\usepackage{amssymb}
\usepackage{bbm}
\usepackage{verbatim}
\usepackage{dsfont}
\usepackage{booktabs}
\usepackage{slashed}

\usepackage{appendix}

\usepackage{epsfig}
\usepackage{color}
\usepackage[table]{xcolor}
\usepackage{graphicx}
\usepackage[]{caption}
\usepackage[listofformat=empty,subrefformat=empty]{subfig}

\usepackage{float}
\usepackage{overpic}

\usepackage{enumerate}
\usepackage{hhline}
\usepackage{multirow}

\usepackage{cite}
\usepackage{xspace}
\usepackage{setspace}

\usepackage[pdftitle={On the weak gravity conjecture in string theory with broken supersymmetry},
  pdfauthor={},
  pdfsubject={},
  bookmarksopen, bookmarksnumbered, bookmarksopenlevel=2, colorlinks=false, linkcolor=blue, citecolor=blue, urlcolor=blue]{hyperref}

\renewcommand{\i}{\text{i}}

\def\be{\begin{equation}}
\def\ee{\end{equation}}
\def\bea{\begin{eqnarray}}
\def\eea{\end{eqnarray}}

\begin{document}

\thispagestyle{empty}

\begin{flushright}
CPHT-RR106.112018\\
IPhT-T18/142
\end{flushright}
\vskip .8 cm
\begin{center}
{\Large {\bf On the weak gravity conjecture in string theory with broken supersymmetry} } \\[12pt]

\bigskip
\bigskip 
{
{\bf{ Quentin Bonnefoy$^{a}$}\footnote{E-mail: quentin.bonnefoy@polytechnique.edu}},  
{\bf{Emilian Dudas$^{a}$}\footnote{E-mail: emilian.dudas@polytechnique.edu}},
{\bf{Severin L\"ust$^{b,a}$}\footnote{E-mail: severin.luest@polytechnique.edu}}
\bigskip} \\[0pt]
\vspace{0.23cm}
{\it $^{a}$ 
CPHT, CNRS, Ecole polytechnique, IP Paris, F-91128 Palaiseau, France\\[01pt] 
\vspace{0.23cm}
\it $^{b}$ 
Institut de Physique Th\'eorique, 
Universit\'e Paris Saclay, CEA, CNRS\\
Orme des Merisiers \\
91191 Gif-sur-Yvette Cedex, France}\\[20pt] 
\bigskip
\end{center}

\begin{abstract}
\noindent

We use type I string models with supersymmetry broken by compactification (\`a la Scherk-Schwarz) in order to test the weak gravity conjecture in the presence of runaway potentials in a perturbative string theory setting.
For a finite value of the supersymmetry breaking radius there is a runaway potential, which is the only possibility if one accepts the non-existence of de Sitter vacua. Although the weak gravity conjecture is valid in the decompactification limit, for fixed values of the radius  we show that there are short-ranged attractive D1 brane-brane interactions. We argue however that at one-loop level the effective tension of the branes decreases and becomes smaller than the effective charge such that there is a long-ranged repulsive force and the weak gravity conjecture is respected. Moreover, for very small $g_s$ we expect a large number of stable bound states to be present.

\end{abstract}

\newpage 
\setcounter{page}{2}
\setcounter{footnote}{0}

{\renewcommand{\baselinestretch}{1.5}


\section{Introduction }
   
Recently several conjectures were put forward constraining the properties of effective quantum field theories which can be consistently UV-completed by a theory of quantum gravity.
These conjectures are usually based on known properties of string theory as well as black hole physics and are often dubbed swampland criteria \cite{Vafa:2005ui}.
Maybe the most prominent of them is the weak gravity conjecture (WGC)  \cite{ArkaniHamed:2006dz}.%
\footnote{For refinements and recent tests of the weak gravity conjecture see \cite{Cheung:2014vva, Heidenreich:2015nta, Kooner:2015rza, Brown:2015iha, Nakayama:2015hga, Harlow:2015lma, Heidenreich:2016aqi, Montero:2016tif, Hebecker:2016dsw, Saraswat:2016eaz, Cottrell:2016bty, Hebecker:2017wsu, Montero:2017yja, Palti:2017elp, Hebecker:2017uix, Fisher:2017dbc, Heidenreich:2017sim, Cheung:2018cwt, Andriolo:2018lvp, Lee:2018urn, Hebecker:2018ofv, Hamada:2018dde, Lee:2018spm, Klaewer:2018yxi}.}
Closely related is the swampland distance conjecture \cite{Ooguri:2006in} and the conjectured absence of non-supersymmetric AdS vacua  \cite{Ooguri:2016pdq}.%
\footnote{For further discussions of these conjectures see also \cite{Klaewer:2016kiy, Andriolo:2018lvp, Heidenreich:2018kpg, Grimm:2018ohb, Blumenhagen:2018nts, Lee:2018urn, Lee:2018spm} and  \cite{Freivogel:2016qwc, Danielsson:2016mtx}.}
Lately, another conjecture \cite{Obied:2018sgi}, often called the de Sitter swampland conjecture, attracted a lot of attention.%
\footnote{Fundamental constraints on the consistency of de Sitter vacua have been previously pointed out in \cite{Conlon:2012tz, Dvali:2017eba}.
After the appearance of possible counter examples \cite{Denef:2018etk, Conlon:2018eyr, Cicoli:2018kdo, Murayama:2018lie, Choi:2018rze, Hamaguchi:2018vtv} the original conjecture has been refined in \cite{Ooguri:2018wrx}, see also \cite{Hebecker:2018vxz}.
Other attempts of refinement were suggested in \cite{Andriot:2018wzk, Dvali:2018fqu, Garg:2018reu, Andriot:2018mav}.
For subsequent discussions in the context of string theory see \cite{Banerjee:2018qey, Aalsma:2018pll, Roupec:2018mbn, Andriot:2018ept, Ghosh:2018fbx, Damian:2018tlf, Dasgupta:2018rtp, Kachru:2018aqn, Kallosh:2018wme, Akrami:2018ylq, Han:2018yrk, Moritz:2018ani, Bena:2018fqc, Kallosh:2018psh, Ellis:2018xdr, Buratti:2018onj, Gautason:2018gln, Olguin-Tejo:2018pfq, Garg:2018zdg, Blaback:2018hdo, Heckman:2018mxl, Blanco-Pillado:2018xyn, Junghans:2018gdb, Emelin:2018igk, Banlaki:2018ayh}.}
This conjecture constrains the scalar potential in a way that forbids the existence of (meta)stable de Sitter vacua in string theory.

In its most common formulation the weak gravity conjecture requires that in the presence of gravity for any gauge interaction there should exist at least one charged particle of mass $m$ and charge $q$ such that (in suitable units) $q \geq m$. This condition can be motivated by the requirement that all charged black holes in the theory should be able to decay without leaving a large number of stable remnants}. Moreover, it makes it impossible to take a smooth limit towards vanishing gauge coupling and therefore ensures that gravity is always the weakest interaction.
These statements allow for a natural generalization for higher-form gauge fields where the charged objects are branes. From the viewpoint of particle-particle (or brane-brane) interactions, the condition $q \geq m$ implies that the electric repulsion between two such particles (or branes) is dominating over their gravitational attraction.
Therefore one could reformulate the weak gravity conjecture as the requirement for the existence of at least one particle or brane for each gauge symmetry such that its effective interaction potential is repulsive.  In the absence of scalar fields this statement is equivalent to the original conjecture \cite{Palti:2017elp}. 
It is the objective of this paper to compute such interaction potentials in explicit string theory models and to test if they are repulsive and if they obey the weak gravity conjecture.

On the other hand, runaway potentials are abundant in string theory and this was considered as a serious phenomenological
problem in the past  \cite{Dine:1985he}.  Motivated by the persistent presence of runaway potentials in string theory, it was also recently conjectured 
in \cite{Agrawal:2018own} that quintessence is maybe the only realistic outcome of a theory of quantum gravity.\footnote{This possibility was entertained earlier in various incarnations. For an earlier
attempt, see e.g. \cite{Damour:2002mi}.}  In this paper we are imposing simultaneously the weak gravity conjecture and 
the existence of a runaway (space) direction in which one field continues to roll.
While in the decompactification limit supersymmetry is restored and the weak gravity conjecture is marginally satisfied, considering the rolling field at a different value generates brane interactions and thus constraints from the point of view of the weak gravity conjecture. 

From a string theory viewpoint, the majority of tests of these conjectures were done in the context of superstring compactifications. On the other hand, supersymmetry breaking generates precisely the ingredients needed for non-trivial tests:  runaway potentials for moduli fields, effective brane-brane interactions and the generation of scalar potentials, potentially interpreted as dark energy. The goal of the present work is to analyze the weak gravity conjecture
in type I string theory with broken supersymmetry. Arguably,  the simplest and best understood way of breaking supersymmetry in string theory is via compactification. This was first proposed at the field-theory (supergravity) level by Scherk and Schwarz  \cite{scherkschwarz},  then applied to heterotic strings \cite{closed_ss}  and then to open strings \cite{Blum:1997cs,open_ss}.  The usual string theory computation of brane-brane interactions \cite{polchinski} can be captured, at large separations $r \gg \sqrt{\alpha'}$, by a field theory computation of tree-level exchange of supergravity massless fields between the branes. 
The setup present however some stringy features that are not fully
captured by a pure field-theory analysis by keeping only the supergravity modes.  Indeed this string theory construction contains, as we review in the next section, odd-winding closed string states with a ``wrong'' GSO projection, which contain the would-be scalar tachyon.
These states do couple to branes and do mediate brane-brane interactions.
Even if in the regime of interest $R \gg \sqrt{\alpha'}$, with $R$ the radius of the Scherk-Schwarz circle, the would-be tachyonic scalar is actually very heavy, its exchange is the main contribution to the brane-brane interactions at long distances that we compute below. 
Due to this feature, we are forced to perform the computations at the string theory level, although the results can be understood to some extent by field-theory arguments.

We use D1 brane interactions as a function of the separation in spacetime as a test of the WGC. We find that at short distances and at one-loop there are attractive forces which have a finite limit where the distance goes to zero, whereas at long distances those attractive forces are exponentially suppressed. Since massive (closed strings) fields do not mediate
long range interactions, our interpretation is that at this order of perturbation theory the branes still have a charge to mass ratio set by the supersymmetric BPS condition. The limit of zero distance suggests that the corresponding self-energy can be interpreted as a negative quantum correction to the tension, which will generate an imbalance between gauge and gravitational forces at higher loops, leading to an effective repulsion at large distances consistent with the WGC. The one-loop attractive forces, unsuppressed at small distances, will induce the formation of a finite number of stable bound states of D1 branes. For very small string coupling, the number of such states can become very large, consistent with the swampland distance conjecture \cite{Ooguri:2006in}.

   The structure of this paper is the following. In Section~\ref{sec:typeIscherkschwarz} we review type I string theory with supersymmetry breaking by compactification. In Section~\ref{sec:runaway} we discuss in more details the resulting runaway potentials. Section~\ref{sec:braneinteractions} deals with the brane-brane interactions at one-loop and their attractive nature, which also allows us to define the quantum corrections to brane tensions.
   In Section~\ref{sec:D1interactions} we notice that D1 branes not only interact among themselves, but they also experience an interaction with the D9/O9 background branes/O-planes. Section~\ref{sec:WGC} contains arguments beyond one-loop, which are needed in order to clarify the fate of the weak gravity conjecture in this setup. The paper ends with some brief conclusions and future directions. 
 \section{Type I strings with Scherk-Schwarz supersymmetry breaking}\label{sec:typeIscherkschwarz}  
   
 Scherk-Schwarz breaking of supersymmetry is the oldest and probably the most popular way of breaking supersymmetry perturbatively in string theory. Since we are interested in brane interactions, moduli potentials and the weak gravity conjecture, the necessary ingredients are present in the type I string and orientifolds \cite{orientifolds}. Vacuum energy and brane-brane interactions are nicely encoded in one-loop string amplitudes: torus and Klein bottle for the propagation of closed strings, and the cylinder and the M\"obius for open strings.  
 In what follows, all string amplitudes below should be multiplied by the factor $1/(4 \pi^2 \alpha')^{d/2}$, where $d$ is the number of noncompact spacetime dimensions. One will add this factor at the end of our computations, in order not to overcharge various formulae. Keeping this in mind, for $9$ non-compact dimensions times a circle of radius $R$ on which the Scherk-Schwarz mechanism is implemented, the one-loop torus amplitude is given by\footnote{For notations and conventions, see \cite{reviews}.}
 \begin{equation}\begin{aligned}
{\cal T} = \int_{\cal F} \frac{d^2 \tau}{\tau_2^{11/2}} \Bigl\{&  (|V_8|^2 + |S_8|^2) \Lambda_{m,2n} - (V_8  {\bar S}_8 +  S_8  {\bar V}_8  ) \Lambda_{m+1/2,2n}    \\
& +   (|O_8|^2 + |C_8|^2) \Lambda_{m,2n+1} - (O_8  {\bar C}_8 +  C_8  {\bar O}_8  ) \Lambda_{m+1/2,2n+1}      \Bigr\}   \frac{1}{|\eta^8|^2} (\tau) \ . \label{ss1}
 \end{aligned}\end{equation}
 where ${\cal F}$ is the fundamental domain of the modular group $\mathrm{SL}(2,\mathbb{Z})$, $V_8,S_8,O_8$ and $C_8$ are $SO(8)$ characters built out of Jacobi theta functions, $\tau$ is the complex parameter of the torus and $\Lambda_{m,n}=\sum_{m,n}q^{\frac{\alpha'}{4}(\frac{m}{R}+\frac{nR}{\alpha'})^2}\overline{q}^{\frac{\alpha'}{4}(\frac{m}{R}-\frac{nR}{\alpha'})^2}$ denotes the one-dimensional lattice of states with Kaluza-Klein (KK) number $m$ and winding number $n$, with $q=e^{2\pi \i\tau}$. Even windings have the familiar action of spacetime fermion number: bosons have the usual KK masses, whereas fermions  have a mass shifted by $1/2R$.  On the other hand, odd winding states have a  different, ``wrong'' GSO projection. In particular, this sector contains a tower of states starting with a 
 scalar (coming from the character $ |O_8|^2$ above) with the lightest mass given by
 \be
 m_O^2 = - \frac{2}{\alpha'} + \frac{R^2}{\alpha'^2}   \ . \label{ss01}
   \ee
 For small radii $R < \sqrt{2 \alpha'}$ this scalar becomes tachyonic, whereas it is very heavy in the opposite limit   $R \gg \sqrt{2 \alpha'}$.  This scalar will be a main actor in the brane-brane interactions at long distances that
 we discuss later on. 
   The Klein bottle amplitude provides the orientifold projection of the closed string sector and is given by
\be
{\cal K} = \frac{1}{2} \int_0^{\infty} \frac{d \tau_2}{\tau_2^{11/2}} \frac{V_8-S_8}{\eta^8} (2 i \tau_2) \sum_m e^{- \alpha' \pi \tau_2 \frac{m^2}{R^2}} \ . \label{ss2}
\ee  
Since it is the same as in the superstring case, it  does not contribute to the vacuum energy and symmetrizes, as usual, the NS-NS sector which comprises the graviton $g_{MN}$ and the dilaton $\Phi$, whereas it antisymmetrizes the RR sector which consists of the two-from $C_2$. 
Consistency of the theory (RR tadpole conditions) requires the introduction of $16$ D9 branes wrapping the circle, which can be endowed with arbitrary Wilson lines \cite{bps} $W_i = \operatorname{diag} (a_i/R, -a_i/R)$, which can 
be interpreted as  D8 brane positions $d_i = 2 \pi a R'$ on the circle after a T-duality, where $R'= \alpha'/R$ is the T-dual radius. 
Notice that in the T-dual interpretation the branes at positions $d_i$ are accompanied by their images under the orientifold projection at $-d_i$. The physically distinct values of the Wilson lines can be chosen to be $0 \leq a_i \leq 1/2$, where the end-points of the interval $a_i=0$ and $a_i=1/2$ correspond to the location of the $O8_{-}$ planes. 
Whereas the T-dual interpretation geometrizes nicely properties of brane spectra and interactions, we should remember that the radius
of the circle is large $R \gg \sqrt{\alpha'}$ in order to avoid the tachyon (and obtained dynamically by the time-evolution). Therefore
the T-dual picture in the supersymmetry breaking radius is not really useful from an effective field theory description, since the T-dual radius is smaller than the string length.

The one-loop open string amplitudes are given by
\begin{align}
  &{\cal A} =    \sum_{i,j=1}^{16} \int_0^{\infty} \frac{d \tau_2}{\tau_2^{11/2}}  \Big[\frac{V_8}{\eta^8}\left(\frac{i \tau_2}{2} \right) (P_{m+a_i-a_j}+ P_{m+a_i+a_j})\nonumber\\ &{\color{white}{\cal A} =    \sum_{i,j=1}^{16} \int_0^{\infty} \frac{d \tau_2}{\tau_2^{11/2}}  \Big[}-\frac{S_8}{\eta^8}\left(\frac{i \tau_2}{2} \right) (P_{m+1/2 + a_i-a_j}  + P_{m+1/2 + a_i+a_j})  \Big]  \ , \\
 &{\cal M} =  -    \sum_{i=1}^{16} \int_0^{\infty} \frac{d \tau_2}{\tau_2^{11/2}}  \left[\frac{V_8}{\eta^8}\left(\frac{i \tau_2}{2} + \frac{1}{2}\right) P_{m+2a_i} - \frac{S_8}{\eta^8}\left(\frac{i \tau_2}{2} + \frac{1}{2}\right) P_{m+1/2+2a_i})   \right]  \ , \nonumber  
   \label{ss3}
\end{align}   
where in this loop channel $V_8$ describes the propagation of (gauge) bosons, whereas $S_8$ that of charged fermions. Moreover,
$P_{m+a_i} = \sum_n e^{- \pi \tau_2 \frac{\alpha' (m+a_i)^2}{R^2}}$ denotes the KK sum of open string states shifted in mass by the Wilson lines. 
The parameter $\tau = i \tau_2/2$ ($\tau = i \tau_2/2 + 1/2$) has the interpretation of the complex parameter of the doubly covering torus for the cylinder (M\"obius) amplitude. 
For generic values of the Wilson lines (brane positions after T-duality), the open string gauge group is $U(1)^{16}$, whereas in their absence it is $SO(32)$.
   The one-loop open string amplitudes have a dual interpretation in terms of tree-level exchange of closed string states between the D-branes (for the cylinder) and between the D-branes and O-planes (for the M\"obius amplitude). The corresponding string amplitudes can be obtained by appropriate modular transformations and are expressed in terms of the length $l$ of the tube describing the tree-level propagation. Doing so, one obtains  
  \bea
 && {\cal \tilde A} =  \frac{2^{-5} R}{\sqrt{\alpha'}}  \sum_{i,j=1}^{16} \int_0^{\infty} dl   \left[   \frac{V_8-S_8}{\eta^8 }(il) 
  \frac{1+(-1)^n}{2} + \frac{O_8-C_8}{\eta^8 }(il) \frac{1-(-1)^n}{2} \right] 
 \nonumber \\
 &&{\color{white}{\cal \tilde A} =  \frac{2^{-5} R}{\sqrt{\alpha'}}  \sum_{i,j=1}^{16} \int_0^{\infty} dl}\times \left[ e^{-2 \pi i n (a_i-a_j)}  +  e^{-2 \pi i n (a_i+a_j)}  \right]  W_n   \ , \\
&& {\cal \tilde M} =  -  \frac{2R}{\sqrt{\alpha'}}   \sum_{i=1}^{16} \int_0^{\infty} d l  \ \frac{V_8  - (-1)^n S_8}{\eta^8}\left(il + \frac{1}{2}\right)  e^{-4 \pi i n a_i}  W_{2n}    \ ,  \nonumber
   \label{ss4}
\eea   
where $W_n = \sum_n e^{- \pi l \frac{n^2 R^2 }{2 \alpha'}}$ denote the (closed string) winding states couplings to the branes-O planes. In (\ref{ss4}) $V_8$ ($S_8$) denote the couplings to the NS-NS (RR) closed string sector, whereas 
$O_8$ ($C_8$) denote the coupling of the odd-winding closed string states with the "wrong" GSO projection. Notice in particular the coupling of the scalar $O_8$ to D9 branes. The corresponding coupling to D1 brane in the next sections  
will play a central role in our analysis.  

Supersymmetry is restored in the large radius limit $R \to \infty$. We therefore expect the dynamics to drive the radius 
to large values. In the region $R \gg \sqrt{\alpha'}$ the would-be tachyonic closed string scalar is very massive and should not be kept in a low-energy effective action. However, due to Jacobi function identities, $V_8=S_8$ and the contribution of the usual NSNS-RR sectors
cancel and the main contribution to D9-D9 brane interactions comes precisely from the exchange of this scalar.  
\section{Scalar potential and runaway vacua} \label{sec:runaway}

The goal of this section is to write explicitly the scalar potential for the radius and the Wilson lines of the D9 branes. 
The scalar potential in string theory is minus the partition function, therefore
\be
V (R,W_i) = - \left( \frac{1}{2} {\cal T} + {\cal K} +  {\cal A} + {\cal M}\right)  \equiv   V_ {\cal T} +  V_ {\cal K}  + V_ {\cal A} +V_ {\cal M}  \ . \label{quin1}
\ee
In the Scherk-Schwarz compactification, supersymmetry is broken by global boundary conditions, which implies that the scalar potential is of field-theory origin in the open part for large radii. It is also of field-theory origin
in the closed string part in the large radius limit.
 The Klein bottle is still supersymmetric and therefore it does not contribute to the scalar potential.      
Supersymmetry is restored in the decompactification limit $R \to \infty$. The  potential can be easily estimated in the regime where effective field theory is valid $R \gg \sqrt{2 \alpha'}$. In this limit, string oscillators
in all amplitudes and winding states in the torus are very heavy and do not contribute.  We can therefore replace the modular functions by their leading contribution, such that 
\bea
 &&  {\cal T} \simeq 128  \int _0^{\infty} \frac{d \tau_2}{\tau_2^{11/2}}   \sum_m \left( e^{- \alpha' \pi \tau_2 \frac{m^2}{R^2}} -   e^{- \alpha' \pi \tau_2 \frac{(m+1/2)^2}{R^2}} \right) 
 \ , \nonumber \\ 
  &&  {\cal A} \simeq 8  \sum_{i,j=1}^{16} \int_0^{\infty} \frac{d \tau_2}{\tau_2^{11/2}}  \left[  P_{m+a_i-a_j}+ P_{m+a_i+a_j} - P_{m+1/2 + a_i-a_j}  - P_{m+1/2 + a_i+a_j}  \right]   \ , \nonumber \\
 && {\cal M} =  -  8  \sum_{i=1}^{16} \int_0^{\infty} \frac{d \tau_2}{\tau_2^{11/2}}  \left[  P_{m+2a_i} -  P_{m+1/2+2a_i}   \right]    \  .   
   \label{quin3}
      \eea
 It is convenient to perform a Poisson resummation of the Kaluza-Klein sums to turn them into winding sums, to get
  \bea
 &&  {\cal T} \simeq 128  \frac{R}{\sqrt{\alpha'}} \int _0^{\infty} \frac{d \tau_2}{\tau_2^{6}}   \sum_n \left[ 1- (-1)^n \right]  e^{-   \frac{\pi n^2 R^2}{ \alpha' \tau_2 }} 
 \ , \nonumber \\ 
  &&  {\cal A} \simeq 8  \frac{R}{\sqrt{\alpha'}} \sum_{i,j=1}^{16} \int_0^{\infty} \frac{d \tau_2}{\tau_2^{6}}   \left[ 1- (-1)^n \right]   \left[  e^{- 2 \pi i (a_i-a_j)n}+  e^{- 2 \pi i (a_i+a_j)n}   \right]  e^{-   \frac{\pi n^2 R^2}{ \alpha' \tau_2 }}   \ , \nonumber \\
 && {\cal M} =  -  8\frac{R}{\sqrt{\alpha'}} \sum_{i=1}^{16} \int_0^{\infty} \frac{d \tau_2}{\tau_2^{6}}    \left[ 1- (-1)^n \right]   e^{- 4 \pi i a_i n}  e^{-   \frac{\pi n^2 R^2}{ \alpha' \tau_2 }}   \ \  .   
   \label{quin4}
 \eea 
As explained at the beginning of Section~\ref{sec:typeIscherkschwarz}, all string amplitudes above should be multiplied by the factor $1/(4 \pi^2 \alpha')^{9/2}$. By including this factor and after a straightforward integration, one gets 
 \bea
 &&  {\cal T}  =  \frac{12}{\pi^{14}} \sum_n \frac{1}{(2n+1)^{10}} \ 
 \frac{1}{R^9}   
 \ , \nonumber \\ 
  &&  {\cal A} \simeq \frac{3}{2 \pi^{14}} \sum_{i,j=1}^{16}  \sum_n \frac{\cos 2 \pi a_i (2n+1) \cos 2 \pi a_j (2n+1)   }{(2n+1)^{10}} \  \frac{1}{R^9}\ , \nonumber \\
 && {\cal M}   \simeq  - \frac{3}{4 \pi^{14}} \sum_{i=1}^{16}  \sum_n \frac{\cos 4 \pi a_i (2n+1)   }{(2n+1)^{10}} \  \frac{1}{R^9}\   \  ,   
   \label{quin5}
 \eea 
  which generate a runaway potential, also typical for quintessence models.  
  Supersymmetry breaking generates therefore runaway scalar potentials, a notoriously well-known fact.  Indeed, all known ways of breaking supersymmetry generate, at some order in the perturbative expansion, a runaway potential which generates a cosmological rolling of the corresponding field towards the runaway infinity. We will not enter here into a phenomenological discussion of such potentials and their viability. The example discussed in this paper is too simple to be viable and is ruled out by time dependence of coupling constants, in particular. More important for our purposes, the vacuum energy is not positive unless one adds Wilson lines. A stability analysis including Wilson lines shows that there are no stable  solutions with positive scalar potential in nine dimensions \cite{steve-herve}. The reason is that in order to increase the value of vacuum energy some D9 branes should be displaced/separated in the T-dual version. However, as discussed in the Appendix, D8 branes (after T-duality) attract each other and such configurations are unstable. In lower dimensions, positive potential with stable brane configurations is possible \cite{steve-herve}  without changing significantly the discussion on the weak gravity conjecture below.  Because of the attractive forces between the T-dual D8 branes, in the next sections we consider the case where there are no Wilson lines on D9 branes. 
    
The formulae above can be generalized easily after compactification to four dimensions. We consider for simplicity a product of circles 
of radii $R_I$, $I = 1, \dots, 6$. In the following we introduce a vectorial notation for the winding numbers ${\bf n} = (n,n_1, \dots, n_5)$ and Wilson lines of the brane $i$, ${\bf a}_i = (a_{i,1}, \dots, a_{i,6})$. 
The vacuum energy, in the large radii limit,  becomes\footnote{We wrote (\ref{comp1}) in the large radii limit. If some dimensions are small $R \sim \sqrt{2 \alpha'}$, $R_I \ll \sqrt{\alpha'}$, 
the expressions (\ref{comp1}) change. First of all, the winding masses along the supersymmetry breaking radius in (\ref{comp1}) come from
the ``wrong'' GSO closed-string sector which have a tachyonic mass contribution and we should really replace $n^2 R^2 \to n^2 R^2 - 2 \alpha'$. If the five additional dimensions are small $R_I \ll \sqrt{\alpha'}$,  only the windings along the supersymmetry breaking radius do contribute to the scalar potential and, whereas for large radii $R_I$ the potential scales as $1/R^9$, for small radii it scales as $1/R^4$. Since our conclusions do not change in this case, in order not to complicate too much the discussion below we consider in most cases the limit of large radii $R, R_I \gg \sqrt{\alpha'}$. } 
 \bea
 &&  {\cal T}  =  \frac{3 \times  2^6 V_6 }{\pi^{9}} \sum_{\bf n}  [1 - (-1)^{n}] \frac{1}{[n^{2} R^2 + n_1^2 R_1^2 + \cdots n_5 R_5^2]^5}   
 \ , \nonumber \\ 
  &&  {\cal A}  \simeq \frac{3 \times 2^3 V_6 }{ \pi^{9}} \sum_{i,j=1}^{16}  \sum_{\bf n}   [1 - (-1)^{n}]  \frac{ \cos (2 \pi {\bf a}_i {\bf n})  \cos (2 \pi {\bf a}_j  {\bf n}) }{ [n^{2} R^2 + n_1^2 R_1^2 + \cdots n_5 R_5^2]^5}  \ , \nonumber \\
 && {\cal M}   \simeq  - \frac{3 \times 2^2 V_6}{4 \pi^{9}} \sum_{i=1}^{16}  \sum_{\bf n}   [1 - (-1)^{n}]   \frac{\cos (4 \pi {\bf a}_i {\bf n}) }{ [n^{2} R^2 + n_1^2 R_1^2 + \cdots n_5 R_5^2]^5}  \  .   
   \label{comp1}
 \eea 
 where in  (\ref{comp1}) $V_6 = \prod_I R_I$. 
  It is now possible to obtain a positive scalar potential for the radii with runaway vacua to infinity. For this, one needs to add Wilson lines and check their stability \cite{steve-herve} .  
 
 For fixed values of the Wilson lines, the 9D effective potential for the radius in the Einstein frame is of the form
  \be     
 {\cal L} = \frac{1}{2\kappa_9^2R^2} (\partial R)^2 - \frac{ce^{\frac{18\phi}{7}}}{R^9}  \  ,  \label{comp2}
 \ee     
where $\phi$ is the dilaton field, $\frac{1}{\kappa_9^2}$ is the nine-dimensional Planck mass and $-\frac{c}{R^9}$ is obtained when summing the three contributions in (\ref{quin5}), according to (\ref{quin1}). After the field redefinition $R = R_0e^{\sigma}$, the radion action becomes
 \be     
 {\cal L} = \frac{1}{2\kappa_9^2} (\partial \sigma)^2 - \frac{ce^{\frac{18\phi}{7}-9\sigma}}{R_0^9}  \  .  \label{comp3}
 \ee 
 
 Supersymmetry is then restored in the  limit $\sigma \to \infty$.   
  The computation above did not take into account the fact that the background spacetime is not static, due to the generated scalar potential.  In particular, the Scherk-Schwarz radius
  is expected to run to infinity in order to restore supersymmetry. This is clearly the case if the potential is positive after compactification, which is possible after adding suitable Wilson lines. 
  Actually, even for negative values of such a scalar potential, the large radius regime, in which supersymmetry breaking is small, is generically reached by cosmological evolution in an expanding universe, as
  shown in \cite{Coudarchet:2018ztz}.
   
\section{Brane interactions and effective brane tensions} \label{sec:braneinteractions}

Type I strings contain charged D9, D5 and D1 branes. They are BPS in the superstring case with their tension equal to the RR charge $T=Q$, which guarantees no interaction between them. There is a subtlety
for the D1-D9 amplitude which does not vanish, but it does so after adding the M\"obius amplitude D1-O9. With supersymmetry breaking turned on, branes start to interact. Our goal is to
analyze this in some detail and to understand the change in the effective tension.  Consider D1 branes wrapping the Scherk-Schwarz circle, charged under the RR two-form $C_2$, which behave like particles 
after compactification, coupling to a gauge field $\int_{S^1} C_2$.   

Let us consider two such D1 branes,  at a distance $r$ in the  transverse coordinates. 
The brane-brane potentials are contained in the cylinder amplitude. Its explicit computation  is very similar to a Casimir vacuum energy.   The interaction is given by
\be
  {\cal A}_{11}  = - \frac{1}{ 2}  \operatorname{Str} \int \frac{dk}{2 \pi} \int_0^{\infty}  \frac{d \tau_2}{\tau_2}  e^{- \pi \alpha' \tau_2 (k^2 + M^2)} \ ,       
   \label{db1}
 \ee 
 with the mass operator given by
 \be
 M^2  \ = \  \frac{1}{\alpha'} N + \frac{(m+a_i-a_j)^2}{R^2} + \frac{r^2}{(2 \pi \alpha')^2}  \ ,       
   \label{db2}
 \ee 
 where $N$ is the number operator for open string oscillators and $W_i= a_i/R$ are open string Wilson lines on the circle.  An explicit computation, similar to the one of D9-D9 brane amplitudes leads to
 the one-loop amplitude  
 \be
 \begin{aligned}
  {\cal A}_{11}  =  \frac{1}{ 2 \pi \sqrt{\alpha'}}   \int_0^{\infty}   \frac{d \tau_2}{\tau_2^{3/2}} e^{- \frac{ \tau_2 r^2}{4 \pi \alpha'}}  \ &\left[  P_{m+a_i-a_j} + P_{m+a_i+a_j} 
  - P_{m+1/2+a_i-a_j}  - P_{m+1/2+ a_i+a_j} \right]\\ &\times\frac{\theta_2^4}{2 \eta^{12}} \left(\frac{i \tau_2}{2}\right) \ .
  \end{aligned}
   \label{db3}
 \ee   
 Written in the (closed string) tree-level channel, the amplitude becomes
\be
{ \tilde  {\cal A}}_{11}  =  \frac{R}{ 4 \pi \alpha'}   \int_0^{\infty}   \frac{d l}{l^4} e^{- \frac{ r^2}{2 \pi \alpha' l}}  \  \left[  1-(-1)^n  \right]   \left[  e^{- 2 \pi i (a_i-a_j)n} + e^{- 2 \pi i (a_i+a_j)n} 
 \right]  \frac{\theta_4^4}{2 \eta^{12}} (i l) e^{- \pi l \frac{n^2 R^2}{2 \alpha'}} \ .       
   \label{db4}
 \ee   
 It is more illuminating to write the  tree-level channel exchange potential in a way which involves an integral over the noncompact momenta of the closed strings exchanged, by using the identity
 \be
 \int_0^{\infty}   \frac{d l}{l^4} e^{- \frac{ r^2}{2 \pi \alpha' l} - \frac{ \pi l}{2}  \alpha' m_n^2} = \frac{\alpha'^3}{8 \pi} \int d^8 k \frac{e^{i {\bf k} {\bf r}}}{k^2 + m_n^2}  \ .    \label{db5}
   \ee
   
Notice that only massive states contribute to the D1-D1 brane interactions. In the region of interest $r, R \gg \sqrt{\alpha'}$ a standard field theory computation  does not capture the string result
 (\ref{db4}).  Indeed,  in the region $r \gg \sqrt{\alpha'}$  the main contribution to the brane-brane interaction comes from the region of a long thin tube $l \rightarrow \infty$ and therefore from the lightest closed string states. 
 However, since the even winding contribution which include the supergravity states vanishes due to a cancellation between the NS-NS and the RR sectors, the main contribution to the interaction comes from odd windings containing
 the would-be tachyon scalar in the closed string spectrum (in character language, $O_8$). 
The D1-D1 brane interactions as seen from the tree-level closed-string (``gravitational'') exchange are given by
 \begin{align}
V_{11}  =  - \frac{  R \alpha'^2}{2\pi^2}\sum_n \int d^8 k  \ e^{i {\bf k} {\bf r}}  \  
 \Bigg[& (1-1) \frac{\cos [4 \pi n a_i]  \cos [4 \pi na_j] }{k^2+ \frac{4n^2 R^2}{\alpha'^2}}  \nonumber \\
 & + \frac{1}{8} \ \frac{\cos [2 \pi (2n+1)a_i]  
 \cos [2 \pi (2n+1)a_j] }{k^2 + \frac{(2n+1)^2 R^2}{\alpha'^2} - \frac{2}{\alpha'}}  \Bigg]  \ .   \label{db6}
 \end{align}
 The contribution of the zero-mode vanishes at one-loop, according to
 our computation, which implies that at one-loop the interaction of D1 branes is still governed by the the BPS tree-level tension and charge $T_1=Q_1$. Indeed, since the one-loop contribution is exclusively mediated by massive states, it is short ranged and therefore cannot be interpreted as coming from an imbalance between the tension and charge of the branes. Actually, since the would-be tachyonic scalar for large radius 
 $R \gg \sqrt{\alpha'}$ is much heavier than the supergravity modes and also heavier than string states,
 one should only keep the terms with $n=0$ and $n=-1$ in the formula above for consistency. 
 
If one fixes the values of the Wilson lines and only considers the dynamics in the dimensions perpendicular to the branes, the short-range one-loop D1-D1  brane interactions are attractive (negative potential) for  coincident position of branes on the circle (zero relative Wilson line $a_i=a_j$) and are repulsive (positive potential)  if the branes are separated, for example if one sits at $a_i=0$ and the second brane sits at the other end of the interval $a_j=1/2$. However, once the dynamics of the Wilson lines is taken into account, one sees
that the potential is such that the only stable point is the attractive one $a_i=a_j$.  
 
 An important output of the computation above is the D1 brane self-energy, obtained by considering a single D1 brane of Wilson line $a$ and setting the spacetime distance ${\bf r}=0$. If the result would be divergent, more care would be needed for its interpretation. However,
 since the result is completely finite and is a contribution localized on the D1 brane worldvolume, it can safely interpreted as a self-energy quantum correction to the brane tension, that we compute here. 
 The interaction is dominated in this case by the integration region $l=0$, which is the UV region of the closed string exchange (IR region of one-loop open strings).  In this case one gets the approximate result
 \be
{ \tilde  {\cal A}}_{11}  =  \frac{8 R}{ \pi \alpha'}   \int_0^{\infty}   d l    \   \sum_n   \cos^2 [2 \pi (2n+1))a_i]  e^{- \pi l \frac{(2n+1)^2 R^2}{2 \alpha'}}        
 =   \frac{16}{ \pi^2 R}   \sum_n   \frac{\cos^2 [2 \pi (2n+1))a_i]} {(2n+1)^2 } \ . 
     \label{db9} 
 \ee  
This amplitude contains brane-brane and brane-image brane interactions. By extracting the brane-brane self-energy, one obtains a correction to the brane tension. One obtains then the one-loop
corrected tension of the D1 brane wrapping the circle, which can be written either as a corrected D1 brane tension or as the mass 
$M_0$ of the wrapped brane on the circle 
\be
T_{1, \rm eff} = T_1 -  \frac{2}{ \pi^3 R^2}  \sum_n   \frac{1} {(2n+1)^2}  \ = \ T_1 -  \frac{1} {2 \pi R^2}  \quad, \quad 
M_0 = 2 \pi R T_{1, \rm eff} 
\ ,  \label{db10}
\ee
where $T_1 = \frac{\sqrt{\pi}}{\sqrt{2}\kappa_{10}} (4 \pi^2 \alpha')$ is the standard type I D1 brane tension. Notice that this one-loop corrected tension is {\it lower} than the tree-level one, due to supersymmetry breaking. Indeed, since $T_1\sim\mathcal{O}(g_s^{-1})$, the correction is of order $\mathcal{O}(g_s)$ with respect to the original value. The tension becomes zero for the special value $R^2 \sim g_s \alpha'$, which is actually in the regime where type I tachyon condenses and the theory is not anymore under control. 
    
   Notice that in a realistic compactification only four spacetime dimensions are noncompact. In this case, the brane-brane potential for $r \gg \sqrt{\alpha'}$ becomes 
 \be
  V_{11} =  - \frac{  R \alpha'^2}{8\pi^2 V_5} \sum_{\bf p} \int d^3 k  \ e^{i {\bf k} {\bf r}}  \  \frac{\cos [2 \pi a_i]  \cos [2 \pi a_j] }{k^2 + m_{\bf p}^2+ \frac{R^2}{\alpha'^2}- \frac{2}{\alpha'}} \ ,  
  \label{db7}
 \ee 
 where $ \sum_{\bf p} $ is the sum over all Kaluza-Klein masses in the five additional internal dimensions.
 
 The result is particularly simple if the five additional dimensions are very small, i.e.~$R_I \ll R,r$, in which case one can neglect the corresponding massive modes contributions. In this limit (and using $R \gg \sqrt{\alpha'}$),
 the total potential energy is well approximated at large distances $r \gg \sqrt{\alpha'}$ by
 \be
 V_{11} \sim   \ -  \   \frac{  R \alpha'^2}{4 V_5}
    \cos [2 \pi a_i]  \cos [2 \pi a_j]   \frac{ e^{-r\sqrt{\frac{R^2}{\alpha'^2}-\frac{2}{\alpha'}}}}{r}    \ .  
  \label{db8}
 \ee
As discussed previously, despite the naive first thought that the potential  (\ref{db8}) is negative for close values of the Wilson lines of the two branes and positive if the branes are well separated on the circle, the only minimum stable configuration is when they are
coincident. 

An important point for the later discussion is that the negative self-energy of D1 branes and the decrease in the effective brane tension also implies that it is energetically favorable to form bound states of D1 branes.
Indeed, let us denote by $V_0 < 0$ the self-energy of one D1 brane. Then one can compare the energy of two configurations.
The first is the energy $E_{N,1}$ of $N$ coincident D1 branes and a single D1 brane at a large distance $r \gg \sqrt{\alpha'}$ from them,
whereas the second is the energy $E_{N+1,0}$ of $N+1$ coincident D1 branes. They are given by
\bea
&&  E_{N,1} = - (N+1) T_1 + (N^2+1) V_0 + O\left(e^{- \frac{r R}{\alpha'}}\right) \ , \nonumber \\
&&  E_{N+1,0} = - (N+1) T_1 + (N+1)^2 V_0   \ .  
  \label{db11}
\eea 
It is then clear that $ E_{N+1,0} < E_{N,1} $ and therefore that the D1 branes tend to form bound states which eventually can lead to the formation of black holes. 

Finally, until now we considered D1 branes wrapping the supersymmetry breaking circle. If on the other hand the D1 branes are perpendicular to the direction of the radius $R$ used for supersymmetry breaking, they do not experience supersymmetry breaking. They will retain therefore the BPS nature at the one-loop level and their interactions will be supersymmetric. 
 \section{D1 interactions with the background D9-O9} \label{sec:D1interactions}       
  
  One natural question is the influence of the background D9-O9 on the potential for the Wilson lines of D1 branes.
 In the type I superstring there is no net interaction between D1 branes and the background D9 branes and O9 planes.\footnote{E.D. thanks Jihad Mourad for a very helpful discussion on this issue.} 
   More precisely, the brane-brane interaction D1- D9 is cancelled by the 
 interaction with the orientifold D1-O9, a consequence of the tadpole cancelation condition and of the BPS properties of type I branes.  
 
 In the case of supersymmetry breaking by compactification, this cancellation does not occur anymore and D1 branes
 feel a net interaction with the background, This generates a potential for the Wilson lines of D1 branes on the circle. In what follows, due to the discussion in Section~\ref{sec:runaway} on the D9 Wilson lines and their attractive nature, we take all T-dual D8 branes to be coincident, i.e.~we introduce no corresponding Wilson lines for the D9 branes. Their addition could change the minima of the D1 positions from this interaction with the background, without changing qualitatively our discussion in the next section concerning brane-brane interactions. As a consequence, as one will check here,  D1 brane interactions with the background D9/O9 tend 
 to stabilize the D1 positions $a_i$ at the origin of the (Scherk-Schwarz) circle.
 The D1-D9 and D1-O9 amplitudes are then given by \cite{dms}  
 \begin{align}
 &{\cal A}_{19}=\frac{32}{2 \pi \sqrt{\alpha'}}  \int_0^{\infty} \frac{d \tau_2}{\tau_2^{3/2}}  \left[   (O_0 S_8 + V_0 C_8) P_{m+a_i} -   (S_0 V_8 + C_0 O_8) P_{m+a_i + 1/2}   \right]  
  \left(\frac{\eta}{\theta_4} \right)^4   \ , \nonumber \\
&{\cal M}_1=\frac{1}{4 \pi \sqrt{\alpha'}}  \int_0^{\infty} \frac{d \tau_2}{\tau_2^{3/2}}  \left[   ({\hat O}_0 {\hat V}_8 - {\hat V}_0 {\hat O}_8) P_{m+2 a_i} -  
  ({\hat S}_0 {\hat S}_8 - {\hat C}_0 {\hat C}_8) P_{m+2 a_i + 1/2}   \right]  
  \left(\frac{2 {\hat \eta}}{{\hat \theta}_2} \right)^4  .   
   \label{19s1}
\end{align}      
In these amplitudes $O_0, V_0,S_0,C_0$ describe the one-loop propagation of open strings scalar, vector and spinors respectively in the two dimensional worldvolume of D1 branes in the light cone formulation, whereas $O_8, V_8,S_8,C_8$ describes the quantum numbers
and degeneracy due to the eight Neumann-Dirichlet coordinates. The corresponding amplitudes in the tree-level / gravitational channel are given by
 \begin{align}
 &{\tilde{\cal A}}_{19} = \frac{32 R }{64 \pi \alpha'}  \int_0^{\infty} dl  \Big[  (V_0 O_8 -  0_0 V_8 + S_0 S_8 - C_0 C_8) e^{- 4 \pi i n a_i} W_{2n}  \nonumber \\ &{\color{white}{\tilde{\cal A}}_{19} = \frac{32 R }{64 \pi \alpha'}  \int_0^{\infty} dl  }+   
 (O_0 O_8 -  V_0 V_8 - S_0 C_8 + C_0 S_8) e^{- 2 \pi i (2n+1) a_i} W_{2n+1}   \Big]  
  \left(\frac{2 \eta}{\theta_2} \right)^4    \ , \nonumber \\
&{\tilde{\cal M}}_1 =  \frac{R}{2 \pi \alpha'}  \int_0^{\infty} dl  \left[ ({\hat O}_0 {\hat V}_8 - {\hat V}_0 {\hat O}_8)  
-  (-1)^n  ({\hat S}_0 {\hat S}_8 - {\hat C}_0 {\hat C}_8)   \right] e^{- 4 \pi i n a_i}W_{2n}  
  \left(\frac{2 {\hat \eta}}{{\hat \theta}_2} \right)^4  .   
   \label{19s2}
\end{align} 
In the tree-level  channel, $V_0,O_0$ denote propagation of NS-NS closed string fields, whereas $S_0,C_0$ denote propagation of RR fields. Notice that there is no net effective interaction in the RR sector exchange,
due to a cancellation between the two terms. This is consistent with the fact that there is no physical RR field to be exchanged between the D1 and D9/O9 sector. The fact that the two amplitudes do not cancel anymore
in the presence of supersymmetry breaking is transparent from the fact that there are odd winding states of the would-be tachyonic (for small radius $R <  \sqrt{2 \alpha'}$) field $O_0 O_8$ in the D1-D9 interaction,
which are not present in the D1-O9 interaction. By summing the two contributions and using identities of Jacobi functions, one finds
\be
 {\tilde{\cal A}}_{19}  + {\tilde{\cal M}}_1 =  \frac{ R }{2 \pi \alpha'}   \int_0^{\infty} dl \ O_0 O_8   \left(\frac{2 \eta}{\theta_2} \right)^4  e^{- 2 \pi i (2n+1) a_i} W_{2n+1}   \ .   
   \label{19s3}  
    \ee
In the limit of interest  $R \gg \sqrt{\alpha'}$, one obtains the leading contribution by taking the limit $l \to 0$ in the string oscillator contributions. By doing so, one finds the final form of the potential
from the 19 sector
\be
V_{19} = - ({\tilde{\cal A}}_{19}  + {\tilde{\cal M}}_1) = -  \frac{ 8 }{\pi^2 R }  \sum_n \frac{\cos [2 \pi  (2n+1) a_i]}{(2n+1)^2}   \ .   
\label{19s4} 
\ee
The minimum of the potential is at $a_i=0$. The result (\ref{19s4}) is valid for large five additional  dimensions $R_I \gg \sqrt{\alpha'}$ and can be understood  as a Casimir field-theory vacuum energy contribution on compact space dimensions. If the five additional dimensions are small $R_I \ll \sqrt{\alpha'}$, (\ref{19s4})
changes and become parametrically of order $(\alpha')^5/V_5 R^6$. This can also be understood by T-dualizing the small dimensions, after which one gets D6 branes wrapping the supersymmetry breaking circle plus five additional large dimensions. The resulting potential energy is of order $V_5'/ R^6$, where $V_5' \gg \alpha'^{5/2}$ is the T-dual volume. This potential is purely field theoretically and can also be understood as a Casimir energy calculation.

This interaction with the background D9 branes/O9-planes seems therefore to favor D1 branes with vanishing Wilson lines. It is unclear and rather implausible to us that this potential energy, localized on the D1 brane but Wilson line/position dependent, should be interpreted as an additional correction to the D1 brane tension. In any case, since it is of the same sign and magnitude as the self energy of the D1 brane, including it or not would not modify the qualitative features of what we discuss next.

 \section{Interactions beyond one-loop  and the weak gravity conjecture} \label{sec:WGC}
 
We consider as in Section~\ref{sec:braneinteractions} two D1 branes separated by a distance $r$ in the three-dimensional noncompact space. Our goal is to estimate their interaction as a function of the distance $r$. We know that at short distances the interaction is attractive and D1 branes tend to accumulate and form bound states. There is no reason to believe that in a perturbative string setup this result would be upset to higher-orders in the perturbative expansion. At large distances however, the one-loop attraction is exponentially damped since the main contribution comes from massive closed-string states. At large distances therefore, potential higher-loop contributions generating massless gravitational (closed string) exchanges would induce infinite-range interactions, which change considerably (and dominate over) the one-loop contribution. This effect can be understood in terms of modifications of the tension and charge of D1 branes, as well as the generation of a dilaton mass, that we now try to include in the interaction potential. All of these modifications are  generated by supersymmetry breaking.

  Let us write the D1-D1 brane interactions in a slightly more general way as a contribution from the zero modes  $V_{11} ^{(0)}$  and contributions from massive states $V_{11} ^{(n)}$. The contribution of the zero-mode $V_{11} ^{(0)}$  vanishes at one-loop, 
  according to our computation in Section~\ref{sec:braneinteractions}. However, since the one-loop contribution comes exclusively from massive states, it is short ranged and therefore any higher-order/loop correction leading to a zero-mode exchange changes
 dramatically the interaction at large distances. We consequently parametrize the zero-mode higher-loop contributions by introducing three parameters:  $T_{1,\rm eff}$ and $Q_{1,\rm eff}$  are the quantum corrected  brane tension and charge, whereas $m_0$ denotes the mass of the dilaton generated by quantum corrections. With these changes in mind, at large distances $r \gg \sqrt{\alpha'}$ where the main contribution comes from the lightest closed string states exchanged between the branes, we arrive at the following expression for the D1-D1 brane interaction
\bea
&& V_{11} =  V_{11} ^{(0)}  +  V_{11} ^{(n)} \,, \quad  {\rm where }  \nonumber \\
&& V_{11} ^{(0)} =   \frac{ R \alpha'^2}{2\pi^2} \int d^8 k  \ e^{i {\bf k} {\bf r}} \  \left[ \frac{Q^2_{1,\rm eff}/Q_1^2}{k^2}  - \frac{T^2_{1,\rm eff}/T_1^2}{4}  \left(  \frac{1}{k^2 + m_0^2}+   \frac{3 }{k^2} \right) \  \right]  \ , \nonumber \\
 &&  V_{11} ^{(n)} =  - \frac{  R \alpha'^2}{8\pi^2} \int d^8 k  \ e^{i {\bf k} {\bf r}}  \  \frac{\cos [2 \pi a_i]  \cos [2 \pi a_j] }{k^2 + \frac{ R^2}{\alpha'^2}- \frac{2}{\alpha'}} \ .   \label{wg1}
 \eea 
 The zero-mode contribution can also be written in terms of the supergravity 10d Planck mass $\kappa_{10}$ as usually done in the literature\footnote{The extra factor of 4 with respect to the usual formula is due to the fact that branes and their images contribute.} \cite{polchinski}
 \be
 V_{11} ^{(0)} =  16 \kappa_{10}^2 \pi R \int \frac{d^8 k}{(2\pi)^8}  \ e^{i {\bf k} {\bf r}} \  \left[ \frac{Q^2_{1,\rm eff}}{k^2}  - \frac{T^2_{1,\rm eff}}{4}  \left(  \frac{1}{k^2 + m_0^2}+   \frac{3 }{k^2} \right) \  \right]  \ .
 \ee
  In (\ref{wg1}),  the corrected tension of the wrapped D1 brane $T_{1,\rm eff}$ is defined  in (\ref{db10}) and the relative factor of 1/4 (3/4) denotes the contribution of the dilaton (graviton).  The one-loop corrected charge  $Q_{1,\rm eff}$ will be discussed below. The massive contributions $V_{11} ^{(n)}$ contain the one-loop computation performed in Section~\ref{sec:braneinteractions}. 
    Notice that in a realistic compactification only four spacetime dimensions are noncompact. In this case, the brane-brane potential becomes 
 \bea
 && V_{11} ^{(0)} =\sum_{\bf p} \frac{16 \kappa_{10}^2 \pi R}{(2\pi)^8V_5}  \int d^3 k  \ e^{i {\bf k} {\bf r}} \  \Biggl[ \frac{Q^2_{1,\rm eff}}{k^2+m_{\bf p}^2}  - \frac{T^2_{1,\rm eff}}{4} \left(  \frac{1}{k^2 + m_{\bf p}^2 + m_0^2}+  
  \frac{3 }{k^2 + m_{\bf p}^2 } \right)  \Biggr]  \ , \nonumber \\ 
 && V_{11} ^{(n)} =  - \frac{  R \alpha'^2}{8\pi^2 V_5} \sum_{\bf p} \int d^3 k  \ e^{i {\bf k} {\bf r}}  \  \frac{\cos [2 \pi a_i]  \cos [2 \pi a_j] }{k^2 + m_{\bf p}^2+ \frac{R^2}{\alpha'^2}- \frac{2}{\alpha'}} \ ,  
  \label{wg2}
 \eea           
 where $ \sum_{\bf p} $ is the sum over all Kaluza-Klein masses in the five additional internal dimensions. As we discussed in the previous sections, in the T-dual version D0 branes energetically prefer to be in the same position and coincident with the D8 branes. Therefore in what 
 follows we can set their position to zero, i.e.~we fix $a_i=0$. Distributing D8 branes on the circle, which would change quantitatively the formulae in this section, raises stability issues and complicates the analysis, without changing  qualitatively the discussion and the conclusions below.
 
 The result is particularly simple if the five additional dimensions are much smaller than $R$ and $r$, in which case one can neglect the contributions from the corresponding massive modes.
 In this limit, it is more transparent to express the  total potential energy in terms of the four-dimensional Planck mass $M_P$, for which the graviton exchange provides the Newton potential
 in terms of the mass $M_0 = 2 \pi R T_{1,\rm eff}$ and the charge $Q_0 = 2 \pi R Q_{1,\rm eff}$ of the wrapped D1 brane. In this way, one gets the approximate potential
 \be
 V_{11} \sim  \  \frac{1}{M_P^2}\left[   \ \frac{ \frac{4}{3} Q_0^2  - M_0^2  -  \frac{1}{3} M_0^2  e^{- m_0 r} }{r}  \ -  \ 
\frac{Q_0^2}{3}   \   \frac{ e^{-r\sqrt{\frac{R^2}{\alpha'^2}-\frac{2}{\alpha'}}}}{r}  \right]  \ .  
  \label{wg3}
 \ee     
 This expression is valid for distances $r \gg \sqrt{\alpha'}$, whereas for shorter distances one expects the one-loop potential
 to be a good approximation, which has a constant limit when $r \to 0$. 
 
 The correction $V_0$ to the D1 brane tension is negative being generated by the massive contributions $V_{11} ^{(n)}$ between the same brane (${\bf r} = 0$). The correction to the charge would, on the other hand, come from a genus $3/2$ computation, which was not yet performed to our knowledge. However, a quantum correction to the RR charge of the brane would be of the form $\int C_2 e^{\phi}$, where $\phi$ is the dilaton. Such a coupling would violate the gauge symmetry of the RR gauge field $C_2$, which seems implausible in perturbation theory. Corrections to the RR field kinetic terms are possible though, and this would generate a renormalization of the RR charge.\footnote{We thank J. Mourad for suggesting this possibility. We also thank I. Antoniadis, G. Bossard, H. Partouche, A. Sagnotti for discussions on this issue.} A similar correction to the dilaton kinetic term should also contribute to the renormalization of the tension. However, such corrections would arise from one loop calculations and would be associated to ${\cal O}(g_s^2)$ corrections. We thus do not expect them to dominate the one-loop contribution to the tension, which is $\mathcal{O}(g_s)$, and therefore
 \be
 T^2_{1,\rm eff} <  Q^2_{1,\rm eff}   \ \Longleftrightarrow 
 \ M_0^2 < Q_0^2 \ .    \label{wg4Bis} 
  \ee
As a consequence, at short distances the potential is attractive whereas it is repulsive at large distances.
If on the contrary the bound \eqref{wg4Bis} was violated in the case of a massless dilaton, i.e.~if $m_0=0$ (or if $M_0^2 > \frac43 Q_0^2$ for $m_0 > 0$), the potential would remain attractive also at large distances. Our perturbative arguments dismiss such a possibility and we conclude that the weak gravity conjecture holds in our setup, and the massless modes exchange which it constrains determines the brane-brane dynamics at large distances.
        
\begin{figure}[htb]
\centering
\includegraphics[scale=0.45]{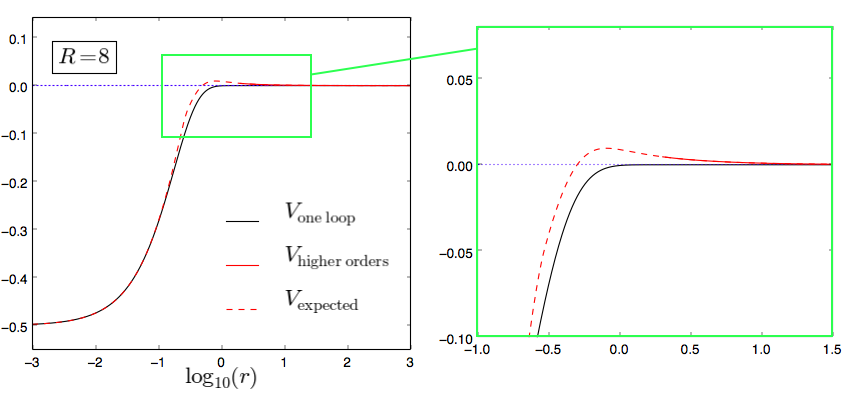}
\caption{The D1-D1 potential as a function of the distance in the transverse space\\
(the potentials and distances are expressed in units of $\alpha'$, we fixed $R=8$, $g_s=0.2$, $V_5\sim1.5^5$ and introduced no Wilson lines for the D1 branes)}
\label{fig:potential}
\end{figure}

Even if \eqref{wg4Bis} holds, the one-loop potential \eqref{db4} between D1 branes is attractive and unsuppressed at small distances, which entertains the possibility that stable bound states, which may be black holes, exist in this theory. Consequently, black holes stability arguments, which are sometimes used in discussions about the WGC, are different in the small and large distance regions. To address this question, one needs to study the regime interpolating between large distances, where higher-order effects dominate and presumably verify the WGC as argued above, and small distances where the one-loop potential induces an attraction. Knowing the $r=0$ value of the potential given in (\ref{db9}) and its asymptotic behaviour (\ref{wg3}), we understand that it reaches a maximal value and has the shape depicted in figure \ref{fig:potential}.

To estimate the location $r_0$ of the maximum, we can use (\ref{wg3}) if $r_0$ is in its validity regime. When $m_0=0$, we obtain 
\be
r_0=-\frac{1}{\sqrt{\frac{R^2}{\alpha'^2}-\frac{2}{\alpha'}}}\left[1+W\left(8\frac{T^2_{1,\rm eff}-T_1^2}{eT_1^2}\right)\right] \approx \frac{\alpha'}{R}\log\left(\frac{R^2}{g_s\alpha'}\right) \ ,
\label{r0}
\ee
where $W$ is the Lambert $W$ function.\footnote{The Lambert $W$ function or product logarithm is defined by $W(x e^{x}) = x$. It has two real branches, here only the lower branch with $W \leq -1$ is relevant.} This expression, obtained from (\ref{wg3}), can be trusted if $r_0\gg \sqrt{\alpha'}$, which can be rewritten as a constraint on the string coupling
\be
g_s\ll \frac{R^3}{\alpha'^{3/2}}e^{-\frac{R}{\sqrt{\alpha'}}} \ .
\label{gSBound}
\ee
In this case, black holes of size smaller than $r_0$ would be stable remnants. Such black holes could be formed from the D1 bound states about which we argued in (\ref{db11}) that their formation is energetically favorable. However, we expect from black hole constructions in string theory that there should only be a finite number of such remnants: from the bound state argument in (\ref{db11}) one can guess that if the number of D1 constituents is large and the bound state size becomes or order $r_0$ or larger, repulsive forces will prevent more D1 branes to bind and therefore larger charge/mass remnants to form. Calculating this finite number of bound states is beyond the scope of this paper, but we could try to estimate it by comparing $r_0$ with the scale at which we expect the D1-branes solutions of supergravity to break down,\footnote{This scale is the one for which the harmonic function $h(r)=1+\frac{R_S}{r}$, which defines the D1-brane solution, starts to deviate significantly from one.} $r_S\sim\frac{N_1g_s\alpha'^3}{V_5}$, where $N_1$ is the number of stacked D1-branes. Using \eqref{r0}, we can derive the following estimate,
\begin{equation}\label{RSr0}
N_{\mathrm{crit}} \equiv N_1 \frac{r_0}{r_S} \approx \frac{1}{g_s} \frac{V_5}{\alpha'^{5/2}} \frac{\alpha'^{1/2}}{R} \log\left(\frac{R^2}{g_s\alpha'}\right) \,,
\end{equation}
where all D1-branes configurations with $N_1<N_{\mathrm{crit}}$ correspond to situations where the attractive force is felt even in the regime where supergravity applies.
In particular, this number becomes small in the decompactifcation limit $R \gg \sqrt{\alpha'}$.

 Furthermore, (\ref{RSr0}) also shows that the smaller $g_s$, the more stable bound states can exist. If $m_0\neq 0$, $r_0$ becomes smaller than (\ref{r0}) and the appearance of such states is slightly suppressed in the limit $g_s \rightarrow 0$, but the behaviour remains qualitatively the same. Such a scaling of $N_\text{crit}$ with $g_s$ seems to be consistent with the swampland distance conjecture.
This conjecture predicts not only the appearance of an infinite tower of massive states as one moves an infinite distance in scalar field space, but also that these states become exponentially light \cite{Ooguri:2006in}.
To test this claim in our setup we compare the masses of the D1-brane bound states with the four-dimensional Planck mass.
Therefore, we should to take the limit $g_s \rightarrow 0$ in such a way such that
\begin{equation}
M^2_P \sim \frac{V_5}{\alpha'^{5/2}} \frac{R}{\alpha'^{1/2}} \frac{1}{\alpha' g_s^2}
\end{equation}
stays constant.
This means we consider
\begin{equation}
g_s, R \rightarrow 0 \qquad \text{such that} \qquad  \frac{R}{g_s^2} = \mathrm{const.}
\end{equation} 
Under this limit we still have $N_{\mathrm{crit}} \rightarrow \infty$ and the bound state masses scale approximately as
\begin{equation}
\frac{M_0}{M_{P}} \sim \left(\frac{\alpha'^{5/2}}{V_5} \frac{R}{\alpha'^{1/2}}\right)^{1/2} \rightarrow 0 \,,
\end{equation}
in agreement with the swampland distance conjecture.
 
On the other hand, when the string coupling increases, $r_0$ decreases and it will eventually not be consistent to use (\ref{wg3}) and (\ref{r0}). Finally, when the supersymmetry breaking (Scherk-Schwarz) radius goes to infinity, as it would be if no further stabilization is added to the dynamics induced by \eqref{comp2} and \eqref{comp3}, the one-loop potential vanishes since supersymmetry is recovered, no attraction nor repulsion remains, and the WGC, as well as the stability of black holes, is marginally retrieved.

 Taking into account the shape of the brane-brane potential, one should clearly also consider the tunneling from large distance $r$ to small ones when discussing the stability of brane configurations. Since we don't have a complete analytic formula, we are unable for the time being to estimate the corresponding tunneling probability. The conditions we derived are therefore necessary but apriori not sufficient to firmly establish the existence of bound states.
      



 \section{Conclusions and perspectives} \label{sec:conclusions}
 
 String theory models with broken supersymmetry usually generate runaway potentials. Such potentials are of exponential type if one canonically normalizes the rolling field and could lead in special cases to quintessence models of dark energy.  
 On the other hand, the breaking of supersymmetry generates at the same time interactions between branes,  which only disappear in the runaway limit. While this in itself respects the weak gravity conjecture 
 at infinity, insisting on the rolling field cosmology could generate violations of it at one-loop, coming from a short-distance attraction generated by massive modes.
Naively one would therefore conclude that in a perturbative and controllable string setting, rolling field dynamics is incompatible with repulsive brane interactions.
It is hence necessary to determine the behavior of the brane-interaction at long distances.
 Since the long-range interaction at one-loop is vanishing due to a cancellation between the massless NS-NS (dilaton and graviton) and the RR exchanges, we believe however that higher-loop corrections are important to settle this issue. We gave qualitative arguments that at higher-loop a repulsive interaction generated by the exchange  of massless states should appear.
At large spatial distances, defined by the parameters $(g_s,R)$, this repulsive interaction dominates over the one-loop (short range) attraction.

The main result of this paper is that in this model, after taking one-loop corrections into account, the effective tension $T_{1,\mathrm{eff}}$ and charge $Q_{1,\mathrm{eff}}$ of D1 branes satisfy the weak gravity bound $Q_{1,\mathrm{eff}} > T_{1,\mathrm{eff}}$.
This is equivalent to a repulsive interaction at long distances as here the aforementioned attractive force is exponentially suppressed.
In the lower dimensional effective theory these D1 branes, wrapped around the Scherk-Schwarz circle, behave as particles charged under a $U(1)$-gauge symmetry with $Q_\mathrm{eff} > M_\mathrm{eff}$.
Overall, this leads to a picture in which the weak gravity conjecture seems to be respected.
To complete our test it would be interesting to compute if supersymmetry breaking induces corrections to the black hole extremality bound as well.

The stability of bound states and black holes is interesting in our setup. The one-loop short-range attraction favors the formation of D1 bound states which can potentially lead to stable black hole remnants. If the string coupling is very small, the attractive region of brane-brane potentials extends up to scales where the effective gravitational theory applies: if $g_s\lesssim \frac{R^3}{\alpha'^{3/2}}e^{-\frac{R}{\sqrt{\alpha'}}}$ (with $R$ the radius of the supersymmetry breaking dimension), a finite number of branes well described by supergravity are sensitive to the attractive potential. This number roughly scales like $\frac{1}{g_s}$, and indicates that in the small $g_s$ limit an increasing quantity of stable bound states is expected to arise.
This behavior agrees with the swampland distance conjecture.

There are a number of open interesting questions that are worth further exploration. It would be interesting to identify string models with broken supersymmetry where the generated moduli potentials and runaway vacua can lead to viable quintessence-like models of dark energy. There are various difficulties for progress into this direction, from generating a small acceleration of the present universe, which is highly nontrivial to achieve in string theory constructions \cite{acceleration}, to the constraints coming from time-dependence of fundamental constants and fifth force experiments. From a more theoretical string theory perspective, it would be interesting to perform 
higher-loop (for instance, genus $3/2$) computations in order to test our result on the quantum corrected brane tension and the absence of renormalization
of the brane charges at lowest order. Whereas supersymmetry breaking should generate, as usual, tadpoles which signal limitations in quantum computations at higher loops, higher-order computations of brane tensions and charges  could be performed by separating two D1 branes in (our) noncompact space, in which case there should be no such problems. It would also be important to investigate stable type I models in lower dimensions with D9 Wilson lines and positive scalar potential in the class of models constructed in \cite{steve-herve} and to investigate the D1 interaction potentials in detail. It would also be very interesting to explore quantum corrections to brane tensions and RR charges in other string models with broken supersymmetry, such as the models with brane supersymmetry breaking \cite{bsb}. 

Finally, we believe it is important to test the other various recent swampland conjectures \cite{Vafa:2005ui, ArkaniHamed:2006dz, Ooguri:2006in, Ooguri:2016pdq, Obied:2018sgi, Ooguri:2018wrx} in explicit perturbative string theory models with broken supersymmetry. Some work along these lines is in progress \cite{bms}.  

\section*{Acknowledgments}

We are grateful to I.~Antoniadis, G.~Bossard and especially I.~Bena, J.~Mourad, H.~Partouche and A.~Sagnotti for enlightening discussions and correspondence.  
The authors acknowledge financial support from the ANR Black-dS-String.
The work of S.L.\ is supported by the ERC Starting Grant 679278 Emergent-BH and the ERC Consolidator Grant 772408 ``Stringlandscape''.


\begin{appendices}

\section{D9 brane interactions}

The D9 brane potentials and interactions are contained in the cylinder vacuum amplitudes. Its explicit computation is very similar to a Casimir vacuum energy and was discussed from the viewpoint of moduli potentials in Sections~\ref{sec:typeIscherkschwarz} and \ref{sec:runaway}.  Consider two D9 branes wrapping the circle, with Wilson lines
$W_i = a_i/R$. After a T-duality, they become D8 branes localized on the circle, of positions $d_i =  2 \pi a_i R'$, where $R'$ is the T-dual radius. In the large radius limit $R \gg \sqrt{\alpha'}$, their interaction is given by
\be
 V_ {\cal A} \simeq - \frac{3}{4 \pi^{14}}  \sum_n \frac{\cos 2 \pi (a_i -a_j)(2n+1)  }{(2n+1)^{10}} \  \frac{1}{R^9}  \ ,       
   \label{bb1}
 \ee 
 The force experienced by the two branes can be computed from
 \be
 F_{ij} = - \frac{\partial V_ {\cal A}}{\partial_{a_{ij}}} =  - \frac{3}{2 \pi^{13}}  \sum_n \frac{\sin 2 \pi (a_i -a_j)(2n+1)  }{(2n+1)^{9}} \  \frac{1}{R^9}  \ .      
   \label{bb2}
  \ee
  This force is attractive, for any value of the radius and any separation $0 \leq a_i-a_j \leq  1/2 $ between the brane positions / Wilson lines. D9 branes are however space-filling objects and therefore cannot be given a separation in spacetime. That is the reason we considered D1 branes to test the WGC. Indeed, D1 branes  wrapped on the supersymmetry breaking circle behave as particles in four-dimensions and can be used to test the gravity as the weakest force hypothesis.  

\end{appendices}
   
\providecommand{\href}[2]{#2}\begingroup\raggedright
\endgroup

\end{document}